\documentclass[sigconf]{acmart}

\usepackage{booktabs} 
\usepackage{draftwatermark}
\SetWatermarkText{DRAFT}
\SetWatermarkScale{1}

\setcopyright{rightsretained}

\acmDOI{}

\acmISBN{}

\acmConference[DS+J]{Data Science + Journalism @ KDD'17}{August 2017}{Halifax, Canada} 
\acmYear{2017}
\copyrightyear{2017}

\acmPrice{15.00}

\begin{document}
\title{Exploring the Ideological Nature of Journalists' Social Networks on Twitter and Associations with News Story Content}

\author{John Wihbey}
\affiliation{%
  \institution{School of Journalism \\ Northeastern University}
  \streetaddress{177 Huntington Ave.}
  \city{Boston} 
  \state{MA} 
}
\email{j.wihbey@northeastern.edu}

\author{Thalita Dias Coleman}
\affiliation{%
  \institution{Network Science Institute \\ Northeastern University}
  \streetaddress{177 Huntington Ave.}
  \city{Boston} 
  \state{MA} 
}
\email{dias.t@husky.neu.edu}

\author{Kenneth Joseph}
\affiliation{%
  \institution{Network Science Institute \\ Northeastern University}
  \streetaddress{177 Huntington Ave.}
  \city{Boston} 
  \state{MA} 
}
\email{k.joseph@northeastern.edu}

\author{David Lazer}
\affiliation{%
  \institution{Network Science Institute \\ Northeastern University}
  \streetaddress{177 Huntington Ave.}
  \city{Boston} 
  \state{MA} 
}
\email{d.lazer@neu.edu}

%

\renewcommand{\shortauthors}{J. Wihbey et al.}
\renewcommand{\shorttitle}{Exploring the Ideological Nature of Journalists' Social Networks}

\begin{abstract}
The present work proposes the use of social media as a tool for better understanding the relationship between a journalists' social network and the content they produce. Specifically, we ask: what is the relationship between the ideological leaning of a journalist's social network on Twitter and the news content he or she produces? Using a novel dataset linking over 500,000 news articles produced by 1,000 journalists at 25 different news outlets, we show a modest correlation between the ideologies of who a journalist follows on Twitter and the content he or she produces. This research can provide the basis for greater self-reflection among media members about how they source their stories and how their own practice may be colored by their online networks. For researchers, the findings furnish a novel and important step in better understanding the construction of media stories and the mechanics of how ideology can play a role in shaping public information. 
\end{abstract}

%
%

\ccsdesc[500]{Applied computing~Sociology}

\keywords{Twitter, journalism, filter bubble, echo chambers, computational social science}

\maketitle

\vspace{-1.5em}

\section{Introduction}

Discussions of news media bias have dominated public and elite discourse in recent years, but analyses of various forms of journalistic bias and subjectivity---whether through framing, agenda-setting, or false balance, or stereotyping, sensationalism, and the exclusion of marginalized communities---have been conducted by communications and media scholars for generations \cite{lazarsfeld_peoples_1968}. In both research and popular discourse, an increasing area of focus has been ideological bias, and how partisan-tinged media may help explain other societal phenomena, such as political polarization \cite{flaxman_filter_2016,mitchell_political_2014,bakshy_exposure_2015}.

The mechanisms that produce such biases in journalistic output, including ideological slant, have been heavily explored at the structural, organizational, and individual levels \cite{patterson_out_2011,entman_black_2001,schudson_sociology_2003,deuze_what_2005}. To date, however, researchers have only been able to study patterns of newsroom work, and its connection to the construction of stories and news agendas, through analytical tools such as surveys, newsroom ethnography, and content analysis.  While these tools are vital, they limit the scale of the analysis and/or are unable to fully capture the social nature of the work of journalism.

In contrast, online social media presents a coarse-grained but large-scale tool for exploring the practice of journalism and the social worlds of those within it.  Social media allows us to measure behavior at granularities more precise than previous methodologies \cite{yardi_dynamic_2010}. Networks of influence can be quantified, and information flows can be understood in relation to embeddeddness within online social communities \cite{romero_interplay_2013}. Social media is important more specifically in the journalistic context because journalists' use of it continues to grow rapidly. Across multiple surveys, journalists report using social media regularly, and microblogging (Twitter) is an especially popular practice that journalists perceive to be instrumental in the profession \cite{santana_tapping_2016}. For example, 54.8\% of journalists report regular use of microblogs \cite{weaver_changes_2016}, while 56.2\% say they frequently find additional information of various kinds on social media and 54.1\% say they commonly find sources there.

In its rapid rise in popularity amongst journalists, social media (and Twitter in particular) not only provides a new platform on which to advance existing practices, they also change the dynamics of news media production and audience-consumer behavior. Consequently, social media presents both a new opportunity to study existing journalistic practices and raises new questions and workflows that demand further inquiry. As journalists spend more time in online communities and as audiences migrate there for news, it is imperative to understand how emerging dynamics are unfolding regarding how news is produced, consumed, and engaged with.

The present work focuses on one specific dimension of social media use by journalists---namely, how the content a journalist consumes and engages with on social media may be ideologically tied to the content he or she produces. In terms of the construction of news stories, two dimensions of journalistic routines might be seen as particularly consequential in news content production and therefore key areas of focus in any study of journalistic social media use: source-related duties; and research-related duties. Journalists report that Twitter's value in news workflow is ``significantly tied to the platform's use for querying followers, conducting research and activities associated with contacting sources'' \cite{santana_tapping_2016}.

It is therefore reasonable to assume that the content a journalist sources on Twitter may impact the content he or she produces. Here, we focus specifically on ideological dimensions of content. Our main research question therefore asks: \emph{what is the relationship between the ideological leaning of a journalist's social network on Twitter and the news content he or she produces?} 

To answer this question, we begin with a novel and unique dataset of journalists linking their Twitter accounts to the articles they write.  We then develop straightforward measures of 1) the ideological leaning of each journalist as determined by who they follow on Twitter and 2) the ideological leaning of each journalist as determined by the articles that they write. After informally evaluating our measures, we then show that there is a modest but significant correlation between them, even when controlling for the outlet a journalist writes for.  This result implies that even within particular outlets (e.g. within the set of political reporters at the \emph{New York Times}), there is a relationship between the ideology of a reporter's online sources and the content they produce offline.

This study is exploratory and cannot assess causation or fully account for confounding variables such as journalists' beats (which may demand deeper embeddedness in specific ideological communities for the purposes of sourcing and research). However, the findings furnish a vital first step in evaluating emerging dynamics in the production of information in an online networked environment. Further, because we can measure the universe of journalists' social networks and correlate this with their off-platform publications, this study affords a unique window into the relationship between online and offline worlds, a larger research domain that remains not yet well understood \cite{bond_61-million-person_2012}.

\section{Related Work}

\subsection{Journalism and Social Media}

The work of journalism is highly governed by routines \cite{shoemaker_mediating_1996}. Emerging work on how social media is being incorporated into these routines suggests that use of platforms like Twitter can become deeply embedded---indeed, ``normalized''---and cross many dimensions of sourcing, information-gathering, and production of stories \cite{lasorsa_normalizing_2012}. Although active participation on social media platforms by individual journalists may have under-appreciated negative costs (e.g. diminished audience impressions of professionalism), most media organizations now actively encourage journalistic activity on such platforms, within limits of organizations' policies \cite{lee_double-edged_2015}.

Journalists also use social media for quasi-marketing functions, a phenomenon that may result in greater responsiveness to feedback from social media. This requires a renegotiation of traditional notions of objectivity and distance from audiences. Problems with the underlying business model of journalism and massive declines in advertising revenue have prompted a diverse set of efforts to engage new audiences and generate online revenue streams. Indeed, journalists may be increasingly ``balancing editorial autonomy and the other norms that have institutionalized journalism, on one hand, and the increasing influence exerted by the audience---perceived to be the key for journalism's survival---on the other'' \cite{tandoc_jr_journalist_2016}.  

Indeed, social media audience responses to stories and the associated analytics have been found to modify newsroom behavior, suggesting that social media can have powerful effects on how newsrooms prioritize information \cite{lee_audience_2014}. Traditionally, journalists have been seen in a hierarchical ``gatekeeping'' role, with power over audiences and sources regarding story selection and prioritization of public agenda items. However, research has found that, because audience interest can now be quantified through online data and newsroom priorities optimized accordingly, ``digital audiences are driving a subsidiary gatekeeping process that picks up where mass media leave off, as they share news items with friends and colleagues'' \cite{lee_audience_2014}.  While the present work focuses on the sources that a journalist can derive from Twitter, and thus does not consider directly the effect of audience feedback on content production, this prior work on the impact of social media interactions and journalistic content provides an expectation for a correlation between online sources and offline content.

A significant amount of more empirical work also exists studying the interrelationships between news media and Twitter. This literature has shown how content in the news and on Twitter coevolve \cite{galle_who_2013,bhattacharya_understanding_2016,kwak_what_2010-1}, and has developed novel methods of linking news articles to tweets about them \cite{tsagkias_linking_2011} and linking both news media and tweets to the same event(s) \cite{wang_mining_2015-1}.  Scholars have also considered how the sharing, and even simply viewing, of news media can expose partisan biases in news consumption \cite{karamshuk_identifying_2016}.

Our efforts differ from this prior work in two ways.  First, prior work mostly focuses on how news content is consumed and spread through social media, or how news agencies choose which content to produce based on social media. In contrast, we consider how the ideological skew of the content consumed by specific journalists on Twitter is associated with the ideology of the content that they generate.  Second, we focus specifically on journalists, rather than the broader Twittersphere.  Here, our efforts are similar to \cite{bagdouri_profession-based_2015}, who develop a method to identify journalists on Twitter.  However, we take a different approach to identifying journalists on Twitter, because we would like to link them to the articles they write. 

\subsection{Inferring Ideological Leanings}

	\subsubsection{From Twitter}
Significant attention has been devoted to extracting political meanings from Twitter, from election prediction (see \cite{beauchamp_predicting_2015} for a nice review) to predicting ideological leanings of users \cite{johnson_identifying_2016,wong_quantifying_2013,golbeck_method_2014,volkova_inferring_2014,ebrahimi_weakly_2016,taddy_measuring_2013-1} to predicting which political party a user is registered for \cite{barbera_less_2016}.  In general, such work shows that Twitter users display their political leanings in a variety of ways, in both the content that they tweet \cite{ebrahimi_weakly_2016,golbeck_method_2014} and whom they follow \cite{barbera_less_2016}. 

The present work focuses on extracting a measure of ideology from the accounts that a journalist follows.  Our efforts are most directly related to those of \cite{barbera_birds_2015}, who develop a Bayesian method for this problem, and \cite{imai_fast_2016-1}, who give a better solution to same Bayesian model. Their work, like ours, relies on the fact that follower relationships are ideologically situated; for example, left-leaning individuals tend to follow more left-leaning accounts.  Their model starts with a set of approximately 4M Twitter accounts as rows of a matrix. Columns of the matrix are then defined by a set of around 400 ``elite political users.''  A cell $i,j$ in the matrix is set to 1 if user $i$ follows elite account $j$.  The authors then use an ideal point model (a Bayesian latent-variable model with a single latent dimension) to infer ideological positions of both users and the elite political users based, roughly, on a decomposition of this matrix.  The authors find that ideologies inferred for elite political users match more traditional measures of ideology by comparing official accounts for congresspeople to traditional measures of partisan ideology.

Our method differs from \cite{barbera_birds_2015} in that we are only interested in inferring ideology of a set of users (journalists).  Consequently, we are willing to provide a significant amount of additional information (i.e. to allow our model to ``know'' existing conventional partisanship measures rather than trying to infer them) to provide as accurate a measure as possible of ideological positions of journalists.

	\subsubsection{From Text}

While ideological detection from Twitter is, necessarily, a relatively new field, the scaling of ideology based on text data has a longer and more varied history (see \cite{monroe_fightinwords:_2008} for a more historical perspective).  In general, approaches to scaling text for ideology require a set of text articles pre-tagged for ideology, for example, a set of books authored by individuals with known ideological leanings  \cite{sim_measuring_2013}.  This text can then be used to determine a set of polarized terms \cite{monroe_fightinwords:_2008} or n-grams \cite{sim_measuring_2013} that are highly polarized.  These terms or n-grams can then be applied to new texts to determine the ideology of the new text, although in some cases the ideological skew and ideology of texts can be jointly inferred \cite{roberts_structural_2013}.

An important limitation of approaches that scale text for ideology in this way, however, is that ideology is often expressed through particular \emph{frames}, where ideology invokes particular sentiments towards a set of bipartisan issues. As others have noted, in such cases keyword-based models like those used here may not fully capture ideological leanings. Instead, scholars have recently developed novel approaches for detecting and extracting frames from text, largely via Bayesian latent variable (i.e. ``topic model like'') approaches \cite{tsur_frame_2015,roberts_structural_2013,card_media_2015}.  While such methods are an important avenue of future work, we here restrict ourselves to a simpler, ideology-based n-gram approach for ease of interpretability and acknowledge the potential limitations of doing so.

\section{Data}

\subsection{Journalist Data}

\begin{table}[t]
\small
  \centering
  \small
  \begin{tabular}{|p{3cm}|p{.5cm}| p{3cm}|p{.5cm}|}
  \hline
  \textbf{Outlet}	& \textbf{N} & \textbf{Outlet}	& \textbf{N}  \\ \hline
	 \multicolumn{2}{|c|}{\textbf{ Heavily Right Leaning}}  &  \multicolumn{2}{|c|}{\textbf{ Heavily Left Leaning}} \\ \hline
RealClearPolitics & 7 & Huffington Post &  68   \\ \hline
Washington Times &  11 &  Vox &  16   \\ \hline
Breitbart & 7  &   Politico &   94  \\ \hline
The Hill & 32 & New Yorker &  13   \\ \hline
National Review & 10 & &  \\ \hline

\multicolumn{2}{|c|}{\textbf{ Right Leaning}}  &  \multicolumn{2}{|c|}{\textbf{ Left Leaning}} \\ \hline
New York Post & 19 & USA TODAY & 45    \\ \hline
Oklahoman & 12 & Bloomberg  & 107  \\ \hline
Tennessean & 9 & BuzzFeed  & 57   \\ \hline
Wall Street Journal & 82 & New York Times & 190     \\ \hline
Arizona Republic & 22 & Washington Post & 143\\ \hline
Florida Times-Union & 13 & &     \\ \hline
Weekly Standard & 9 & & \\ \hline
Dallas Morning News & 25  & & \\ \hline
New York Daily News & 25 & & \\ \hline
Orange County Register & 12  & & \\ \hline
Las Vegas Review-Journal & 14 & & \\ \hline
  \end{tabular}
\caption{A list of the 25 news outlets used in the present work. The columns labeled $N$ give the number of journalists from each outlet that met our criteria for inclusion.  Outlets are broken up by traditional expectations of (heavily or not) left/right leaning content}
  \label{tab:contexts}
  \vspace{-2.5em}
\end{table}

We began with a set of 25 news outlets chosen with the goal of finding a relatively balanced sample of news sources from across the political spectrum. An initial set of outlets was chosen from a Pew Research Center study \cite{mitchell_political_2014-1} that looked into the ideological configuration of the media audience. From the Pew list we selected outlets whose media format was either newspaper, magazine or blogs. Finding the resulting list to be skewed toward liberal news outlets, we applied our domain knowledge to add more right-wing news outlets to the sample. 

For each outlet, we then extracted all journalists associated with the outlet according to MuckRack\footnote{\url{http://www.muckrack.com}}---an online platform for public relations professionals interested in tracking news media. Journalists who were identified by MuckRack as writers of Arts and Entertainment, Food and Dining, or other non-political sections were excluded from our sample. All remaining journalists were assumed to be political journalists, insofar as they covered topics that had a public affairs dimension.

For each political journalist, we then obtained both their Twitter handle and a list of their published articles from MuckRack. Journalists with fewer than 100 published articles were excluded from the sample, as well as those that did not have an accessible Twitter account.  The set of 25 outlets studied, along with the number of journalists we considered from each outlet, is shown in Table~\ref{tab:contexts}.  Additionally, we give a loose categorization of the news article into those that are heavily or slightly right or left-leaning. After cleaning and filtering steps described, we were left with a total of 1,047 journalists, who wrote a combined set of 502,340 articles.


\subsection{Congressional Data}

We use three datasets derived from members of Congress. First, in order to determine the political affiliation and official social media accounts for members of Congress, we leverage a database linking congresspeople to their official Twitter handles\footnote{\url{https://github.com/unitedstates/congress-legislators}}.  Second, as we discuss below, it is also useful to have a measure of the degree to which each congressperson leans left or right. For this, we take \emph{DW Nominate scores} \cite{poole_d-nominate_2001} of each candidate provided by GovTrack\footnote{\url{https://www.govtrack.us/about/analysis}}. DW Nominate scores are a standard measure in political science of the position of each congressperson on a [0,1] interval (which we translate to [-.5,.5]). They are measured by extracting a latent ideological dimension derived from congressional voting patterns.

Finally, in order to scale newspaper articles on an ideological scale, we rely on public statements made by congresspeople to determine ideologically situated language.  We extract a set of the last 150,000 public statements (dating back to mid-2015) made by members of Congress captured on the website VoteSmart.org\footnote{\url{http://votesmart.org}}.  Prior work on the same dataset has shown this corpus to be useful for understanding politicized language \cite{tsur_frame_2015}.

\section{Methodology}

Our focus is on understanding the ideological relationship between journalists' social network and the news they generate. In this section, we describe how we construct measures of journalists' ideology \emph{as represented by} 1) who they follow on Twitter and 2) the news articles they write. These measures serve as proxies, respectively, for the ideology of journalists \emph{as represented by} their network ties and their news outputs. We then take these proxies to study the underlying phenomena of interest.

\subsection{Ideology of Twitter networks}

Perhaps the simplest method to determining ideology of a Twitter user from their following relations, as has been done elsewhere \cite{cohen_classifying_2013}, is to count the number of Democrat vs. Republican congresspeople the user follows. However, there are many Twitter accounts that do not follow congresspeople. More importantly, many accounts that users follow that are not congresspeople are still indicative of political ideology. This is particularly true of those on the far-right of the political spectrum. We therefore develop a method that leverages known information about congresspeople only indirectly, through a more complete picture of the accounts an individual follows. We first project journalists, the congressional accounts they follow, and other accounts that are strong indicators of political preferences, into a shared latent space. We then  use known ideology scores of congressional accounts to project points in this latent space onto an ideological dimension.

We first follow \cite{barbera_birds_2015} and construct a matrix of users for whom we wish to infer ideology based on follower relationships (rows) and the \emph{elite} they follow (columns). We define elite accounts as any account for which we have a DW-Nominate score (n=502) or any account followed by more than 2\% of all rows (n=2,549).  The rows of our matrix are defined by two distinct sets of Twitter users. First, we include the full set of journalists we are interested in studying.  Second, we include a set of 12,001 highly \emph{politically active} Twitter users. We determine the set of politically active users by first linking Twitter accounts to public voter registration data, as is done in \cite{barbera_less_2016}.  We consider an account to be politically active if it is both a) a registered Democrat or Republican and b) follows at least 3 congressional accounts.  Politically active accounts are useful for two reasons. First, they can be used to ensure that our methodology correctly exposes ideological differences in following relations between left and right-leaning accounts.  Second, they encourage the latent space representation of the matrix to include a strong ideological component.  

Having constructed this matrix, we then use singular value decomposition (SVD) to project both the rows and the columns into a shared low-dimensional latent space. We follow common practice in the word embedding literature and first transform the matrix to represent positive pointwise mutual information (PPMI) \cite{levy_improving_2015-1}.  We then run SVD with $k=5$ latent dimensions---note that qualitative results presented are robust to varying $k$ beyond this, but that five latent dimensions was found to be enough for the present work.

We then construct a regression model that determines how ideology is represented in this low-dimensional space. We regress DW-Nominate scores of congressional accounts (a subset of the matrix columns) on the 5-dimensional representation of these accounts produced by the prior step. The regression model we train is a generalized additive model (GAM) \cite{wood_generalized_2006}, and explains 83\% of the variance.  Finally, because the rows and columns of the matrix are represented in the same latent space, we can use the parameters of this regression model to project the rows onto this same ideological spectrum. This results in a final score for each journalist of the ideological leanings of the accounts they follow, as well as an ideological score for each of the politically active accounts.

With respect to evaluation, we are interested in understanding how well our model captures ideological structure in the rows of our matrix.  In Section~\ref{sec:results}, we therefore assess how well our model captures party registration differences in politically active accounts, and how well it matches a well-known ideological score of 500 websites, which include domains for many outlets studied here, derived from Facebook data \cite{bakshy_exposure_2015}.

\vspace{-2mm}
\subsection{Ideology from News Content}

Our methodology for determining ideological slant of journalists based on the articles they write is completed in two steps that lean heavily on the methodology proposed in \cite{monroe_fightinwords:_2008}.  We first extract phrases indicative of a left or right leaning ideology, and then we score journalists based on the number of times they express these terms in their articles. We discuss each of these pieces in more detail below.

\vspace{-1.5mm}
\subsubsection{Finding Ideologically Informative Terms}

Our first step is to construct a set of phrases (here, terms) that are representative of a strong left or right leaning ideology.  For this purpose, we use the corpus of 150,000 congressional public statements derived from VoteSmart.  The terms extracted from each statement are obtained using the recently developed NPFST algorithm \cite{handler_bag_2016}, which extracts multiword phrases via part-of-speech patterns.  Table~\ref{tab:words} below displays some of the terms extracted.

After extracting terms, we construct a single bag-of-words representation for each congressperson from all public statements written individually by him or her. Following \cite{monroe_fightinwords:_2008}, we then score each term using a smoothed log-odds ratio of the frequency with which the term was used by Democrats versus Republicans. Mathematically, the score we extracted for a term $t$, $s(t)$, is calculated using Equation~\ref{eq:lo}, where $y_t^D$ ($y_t^R$) represents the number of Democrats (Republicans) who used the term $t$ at least once, $\lambda$ is a smoothing parameter\footnote{Or equivalently, a Dirichlet prior; $\lambda=1$ for all experiments here}, $|T|$ is the number of terms considered, and $|D|$ and $|R|$ are the number of Democrats and Republicans considered, respectively ($|D|=229$, $|R|= 276$, $|T|=642,990$).
\begin{equation} \label{eq:lo}
	s(t) = \log(\frac{y_t^R + \lambda}{|R| + (|T|-1)\lambda - y_t^R}) - \log(\frac{y_t^D + \lambda}{|D| + (|T|-1)\lambda - y_t^D})
\end{equation}

Equation~\ref{eq:lo} discounts the frequency with which terms were used by each congressperson. While in the future there may be useful ways to incorporate these values, we chose to construct our measure based solely on the number of congresspeople who expressed each term in order to construct a set of terms with broad usage across the entire party, rather than by any particular subset within a party (e.g. the Freedom caucus).

After scoring all terms, we extract the top 100 most left-leaning (negative) and right-leaning (positive) terms as measured by $s(t)$. We then follow \cite{sim_measuring_2013} and use domain expertise to narrow these sets down into those most indicative of ideology. We found this step to be necessary due to the important differences between congressional statements and news articles. Specifically, congressional statements tended to be more polarized in what kinds of news they would discuss (e.g. Democrats were much more likely to discuss mass shootings), while news agencies were more likely to discuss all forms of news but to have a particular slant towards them.  While this points to future work in modeling both slant and ideological term selection, we found that domain knowledge could determine concepts that were expected to be indicative of ideological leaning if used at all, regardless of slant.

From the top 100 terms found via scoring using Equation~\ref{eq:lo}, we extracted 57 right-leaning terms and 43 left leaning ones. In order to ensure we had the same number of terms for each side, we then filled in an additional 14 left-leaning terms from the following 50 most highly left-leaning terms.  The full set of terms used, along with the code and data necessary to reproduce analyses, are available as part of the code release at the link given above.

\subsubsection{Determining Ideology of Journalists}

Given sets of left and right-leaning terms from above, we then determine an ideological score for each journalist, $s(j)$, via the (smoothed) log-odds of the journalist using a left-leaning term as opposed to a right-leaning term across all of his or her articles:
\begin{equation}
	s(j) = \log(\frac{ y_j^R +1}{y_j^D+1})
\end{equation}

For analyses involving journalists scored for ideology on text, we restrict ourselves to the 648 journalists who wrote at least ten articles with more than 200 words and who expressed at least one ideological term in their writing.

\section{Results}\label{sec:results}

Here, we first provide informal evaluations of our methods for Twitter and news content. We then consider our primary research question in Section~\ref{sec:res_both}.
\subsection{Twitter Ideology}

\begin{figure}
	\begin{tabular}{cc}
		\includegraphics[width=.23\textwidth]{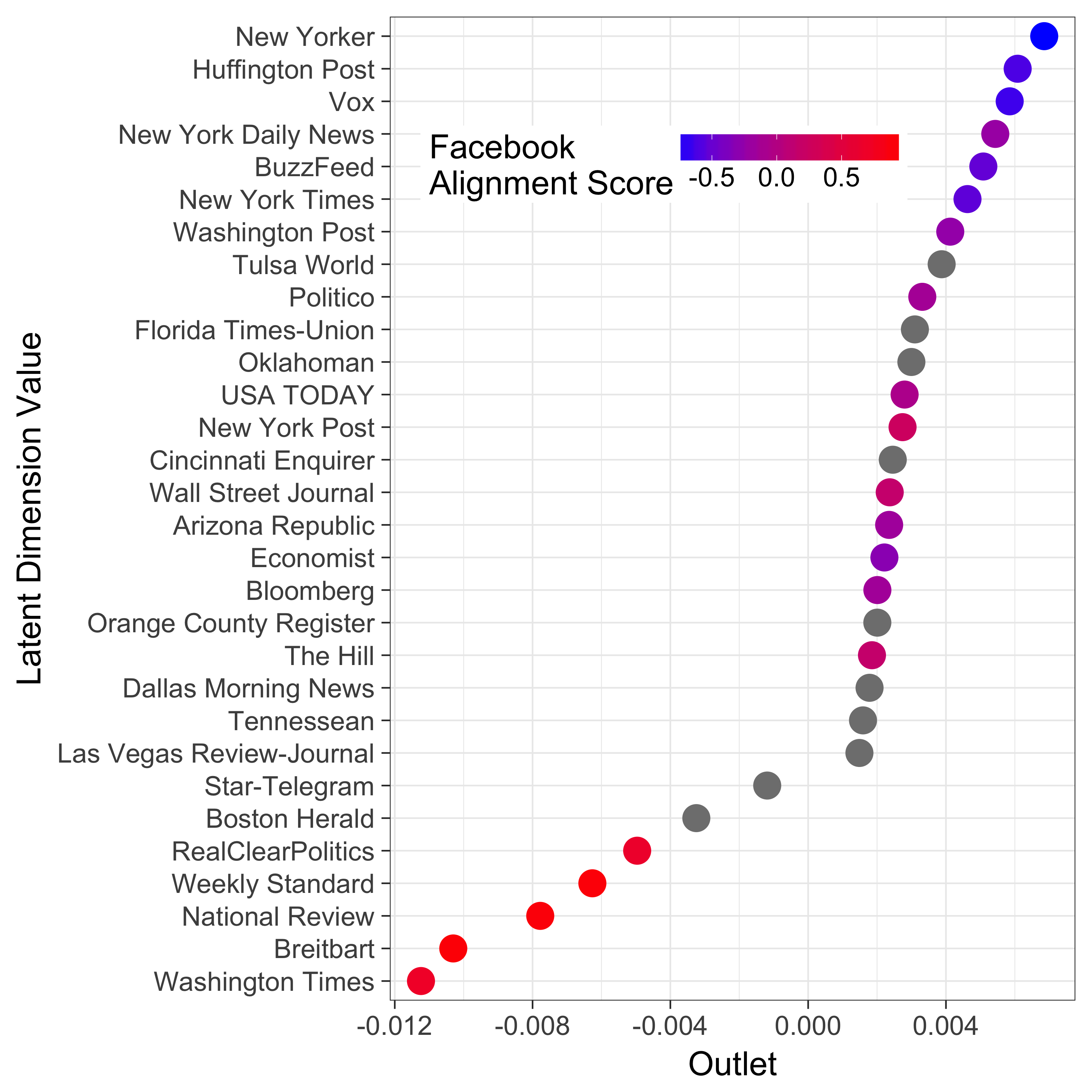} &
		\includegraphics[width=.23\textwidth]{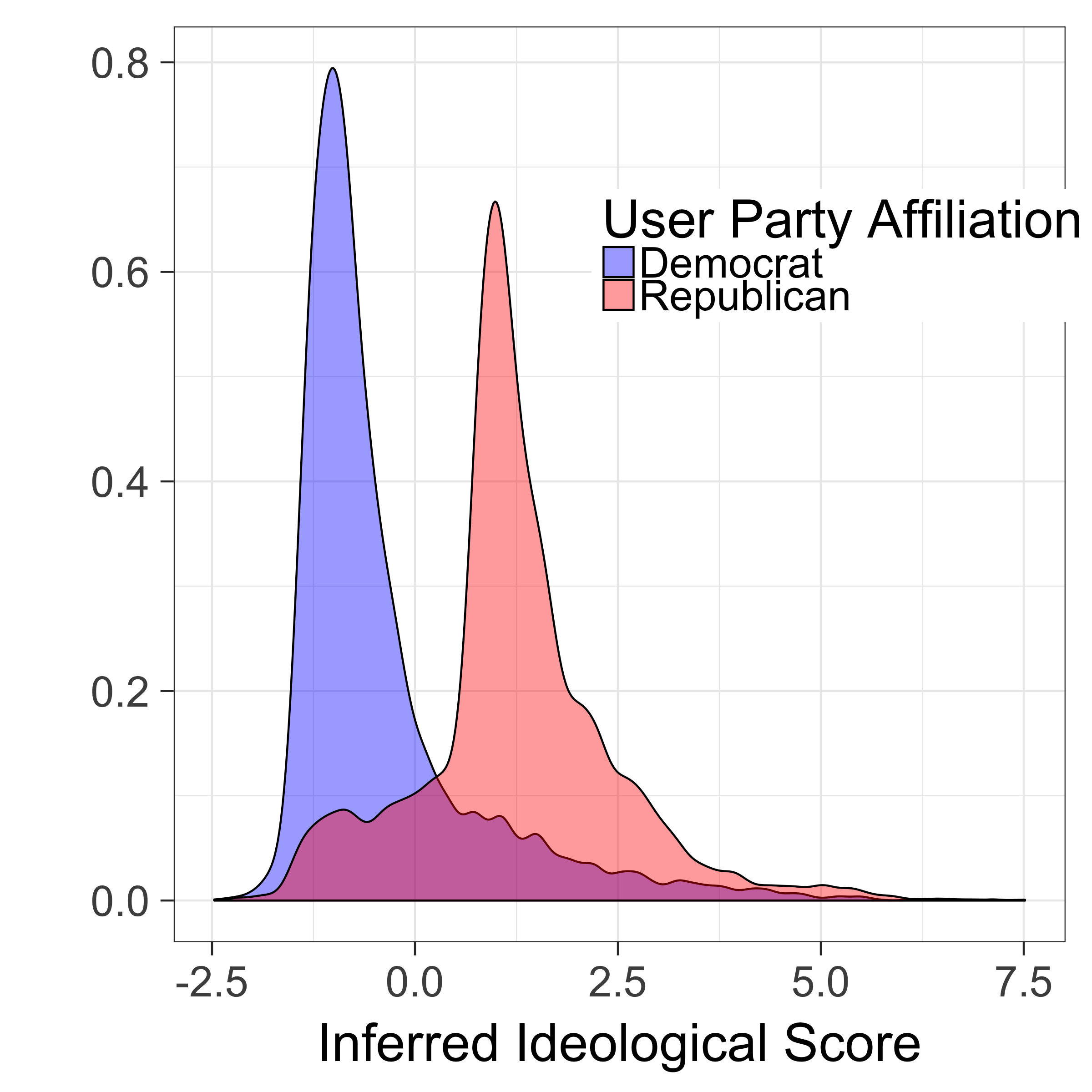} \\
		(a) & (b) \\ 
	\end{tabular}
	\caption{a) On the x-axis, the average weight of all journalists for the given news organization (y-axis) on the latent dimension most correlated with ideology in the Twitter data. Points are colored by ideology of the agencies website by \cite{bakshy_exposure_2015} (blue is more left-leaning, red more right-leaning). Where no estimate is available, the point is colored grey.  b) Distribution of predicted ideology scores for registered Democrats and registered Republicans in our sample.}
	\label{fig:diagnostic_twitter}
	  \vspace{-1.5em}
\end{figure}

Figures~\ref{fig:diagnostic_twitter}a) and b) present two informal evaluations of our method for extracting ideology from the accounts a Twitter user follows. Each provides evidence that the method we use does a reasonable job extracting the desired quantity.  

In Figure~\ref{fig:diagnostic_twitter}a), the y-axis defines the set of news outlets considered in this study, and the x-axis gives the average score of all journalists for that outlet on the latent dimension our regression model identifies as the most correlated with ideology.  Points for each outlet are colored with an estimate of the ideology of that outlet's audience from a well-known prior work on partisan Facebook sharing behavior \cite{bakshy_exposure_2015}.  Outlets not measured in the prior study are colored grey.  Figure a) shows a strong relationship (Pearson correlation of .92) between ideological measurements of previous work and the proposed method.  Further, we are able to score several new outlets on the partisan scale in a way that matches prior intuitions of ideological leanings.

In Figure~\ref{fig:diagnostic_twitter}b), we plot the distribution of ideology scores for the 12,0001 politically active users we infer ideology for, split by the set of accounts registered as Democrats and those registered as Republicans. Again, a clear partisan division, well-known to exist on Twitter (e.g. \cite{barbera_birds_2015}), emerges from our analysis, giving us further confidence in the quality of our measurements.

Finally, it is also interesting to explore how the method ranks specific journalists ideologically based on who they follow. We found that journalists with the most heavily right-leaning followerships are at traditionally right-leaning outlets such as the \emph{Washington Times}, \emph{Breitbart} and \emph{The Hill}. Similarly, journalists following the most left-leaning accounts tend to be from left-leaning outlets. However, there are interesting exceptions. For example, among the journalists following the most right-leaning accounts are Eliana Johnson (\emph{Politico}), Jeremy Peters (\emph{New York Times}) and James Hohmann (\emph{Washington Post}).  

While any evaluation of these results is largely subjective, we note that they were produced by data scientists with no knowledge of these journalists. When evaluated by paper authors with a journalist bent, however, it was acknowledged that Jeremy Peters frequently covers Republicans, that Eliana Johnston is a well-known conservative writer and that James Hohmann now regularly writes analytical articles about the Trump administration. These observations provide further face validity for the method and generated new avenues of understanding of how ideology of a reporters content intersects with their online source environment.

\subsection{News Content Ideology}\label{sec:news}

\begin{table}
\small
\begin{tabular}{|p{.21\textwidth}|p{.23\textwidth}|} \hline
\textbf{Top Left-Leaning Terms} & \textbf{Top Right-Leaning Terms} \\  \hline
                 \textbf{lgbt} & \textbf{bureaucrats} \\ \hline
                  voting rights &  obamacare \\ \hline
                  lgbt community &  \textbf{overreach} \\ \hline
         gun safety & \textbf{waters of the united} \\ \hline
\textbf{comprehensive immigration reform} &  waters of the united states \\ \hline
               \textbf{equal pay} & state sponsor \\ \hline
      comprehensive immigration & \textbf{burdensome regulations} \\ \hline
         \textbf{voting rights act} &      \textbf{deal with iran} \\ \hline
          minimum wage & reconciliation act \\ \hline
                     mass shooting & \textbf{mandates} \\ \hline
         \textbf{marriage equality} & \textbf{executive overreach} \\ \hline
        \textbf{sexual orientation} & \textbf{separation of powers} \\ \hline
     \textbf{violence prevention} &  nuclear deal with iran \\ \hline
                   \textbf{equal work} & \textbf{detainees} \\ \hline
          orientation & \textbf{illegal immigrants} \\ \hline
  gun violence & state sponsor of terrorism \\ \hline
        \textbf{bigotry} & \textbf{sponsor of terrorism} \\ \hline
\end{tabular}
\caption{The top 15 left and right-leaning terms as determined by our scaling method.  Terms in bold were selected for the final list of 114 terms (57 left-leaning and 57 right-leaning) to scale news articles} 
\label{tab:words}
\vspace{-2.5em}
\end{table}

Table~\ref{tab:words} presents the 15 words identified as being the most left-leaning (left column) and right leaning (right column) as measured by $s(t)$, and give face validation for our approach. Table~\ref{tab:words} also shows the utility of using a phrase-based generator \cite{handler_bag_2016}---n-grams like ``separation of powers'' and ``voting rights act'' are much more useful and descriptive measures of partisan views and topical foci than, e.g., ``separation'' or ``voting''. 

\begin{figure}
	\includegraphics[width=.4\textwidth]{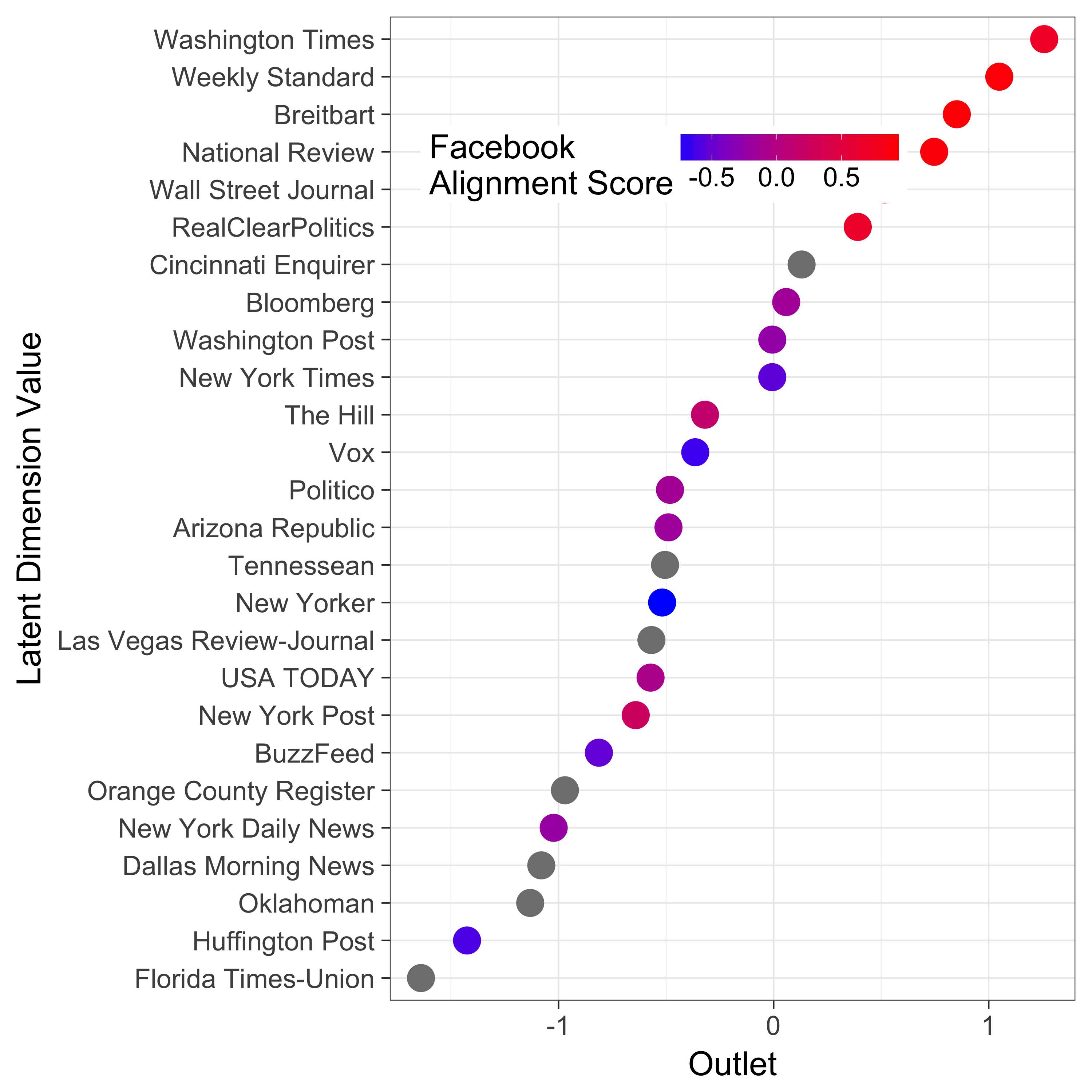} 
	\caption{Average ideology score of all journalists (x-axis) for a news organization (y-axis) as determined by the articles those journalists wrote. Points are colored by ideology of the agencies' website as given in \cite{bakshy_exposure_2015} (blue is left-leaning, red is right-leaning). Grey indicates no such estimate exists.}
	\label{fig:diagnostic_news}
	\vspace{-1.5em}
\end{figure}

In addition to assessing the quality of the partisan terms extracted from congressional statements, we consider how well these terms can be used to identify partisanship of journalistic content.  Figure~\ref{fig:diagnostic_news} shows, as with the Twitter measure, that far-right leaning outlets are clearly separable from the other news outlets studied using our measure. 
However, we also see that across all other outlets, both the prior measure of partisanship (from \cite{bakshy_exposure_2015}) and our expert intuitions given in Table~\ref{tab:contexts} are more weakly correlated with ideological content of news articles than with Twitter networks.  As noted, this may be driven in part by simplistic method we use to identify ideology, which does not do enough to identify framing of general issues. Additionally, it is possible that the recent shifts in the political landscape of U.S. politics led more conservative-leaning outlets to skew more liberally on politics (though perhaps not, for example, on economics, which we do not study here).

Despite these limitations, our approach to identifying ideology of news articles is grounded in a significant literature that suggests partisan terms can be used to identify at least some level of partisanship in writing \cite{monroe_fightinwords:_2008}. Further, results here do align outlets at political extremes correctly relative to each other (e.g. Huffington Post and Vox on the left, Breitbart and National Review on the right). Consequently, we proceed to our main research question with confidence that our approach is able to identify some level of partisan ideology from text.

\subsection{Comparing Twitter and News}\label{sec:res_both}

\begin{figure}
	\includegraphics[width=.45\textwidth]{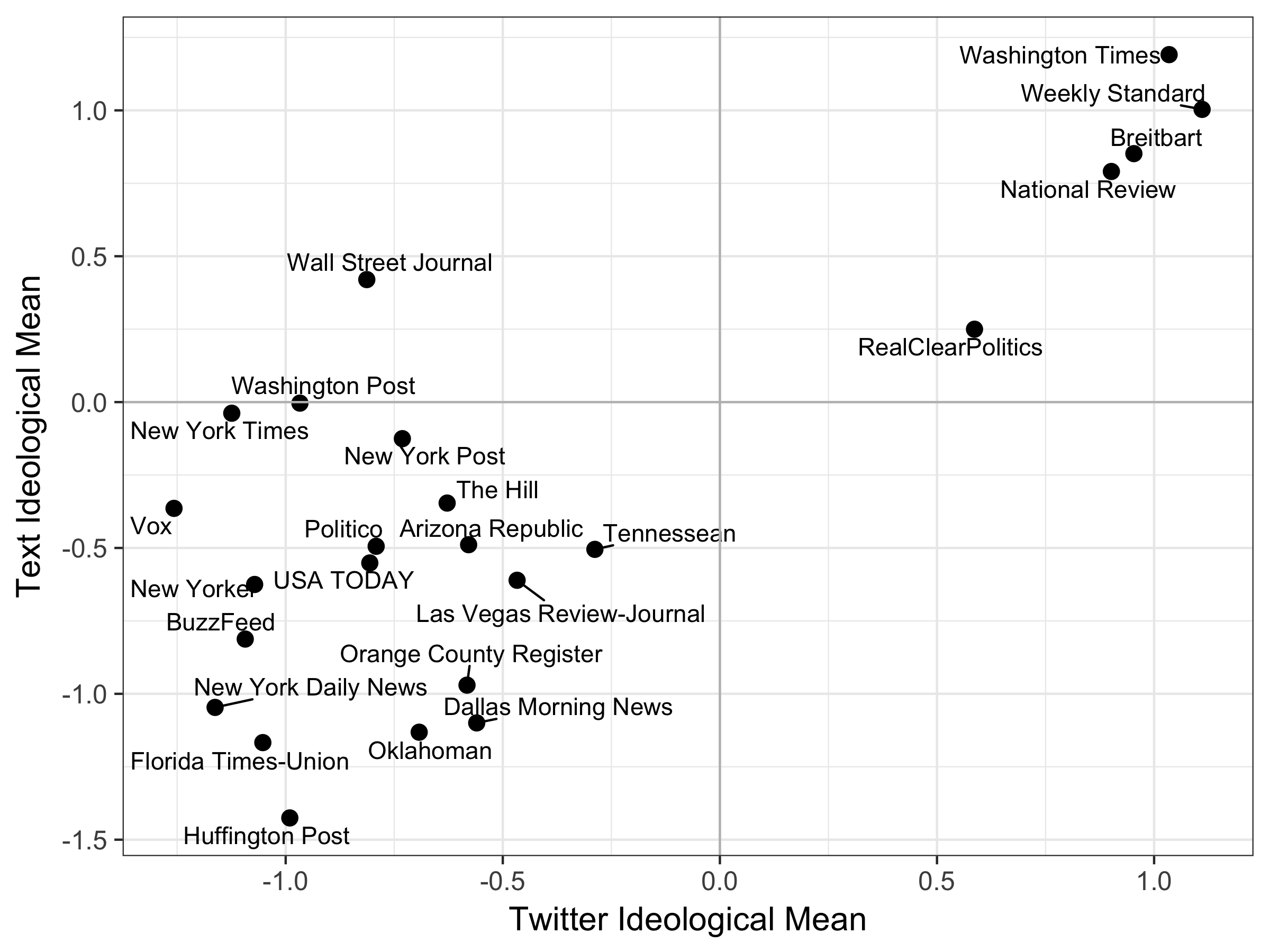}
	\caption{Each point is a news outlet. On the y-axis, the average ideological rating of all journalists at the outlet as estimated by the articles they wrote.  On the y-axis, the average ideological rating of all journalists at the outlet as determined by whom they follow on Twitter}
	\label{fig:scatter_outlet}
\vspace{-1.3em}
\end{figure}

Figure~\ref{fig:scatter_outlet} compares the mean Twitter-based (x-axis) and news content-based (y-axis) ideological scores of journalists for each outlet, and shows a clear positive relationship between the two---the more left-leaning accounts the journalists for a given outlet follow, the more left-leaning their content tends to be.

Beyond this general observation, two points are of note in Figure~\ref{fig:scatter_outlet}.  First, the clear break between the heavily right-leaning outlets (in the top right) and all other media outlets in both Twitter and news content provides further evidence of the rapid disintegration of the ``center-right'' \cite{benkler_study:_2017-1}, leading to a new level of extreme right-wing philosophy impacting political dialog.  Second, we see that the three most established news agencies we study---the \emph{New York Times}, \emph{Washington Post} and the \emph{Wall Street Journal}---have largely left-leaning Twitter networks but produce fairly ideologically neutral or even fairly conservative content. This result suggests that the ``left-leaning bias'' these media sources have been accused of is likely due not only to how ideological content is framed but also to  a purely social ``filter bubble'' that segregates journalists at these outlets from more right-leaning communities. 

\begin{figure}
	\includegraphics[width=.4\textwidth]{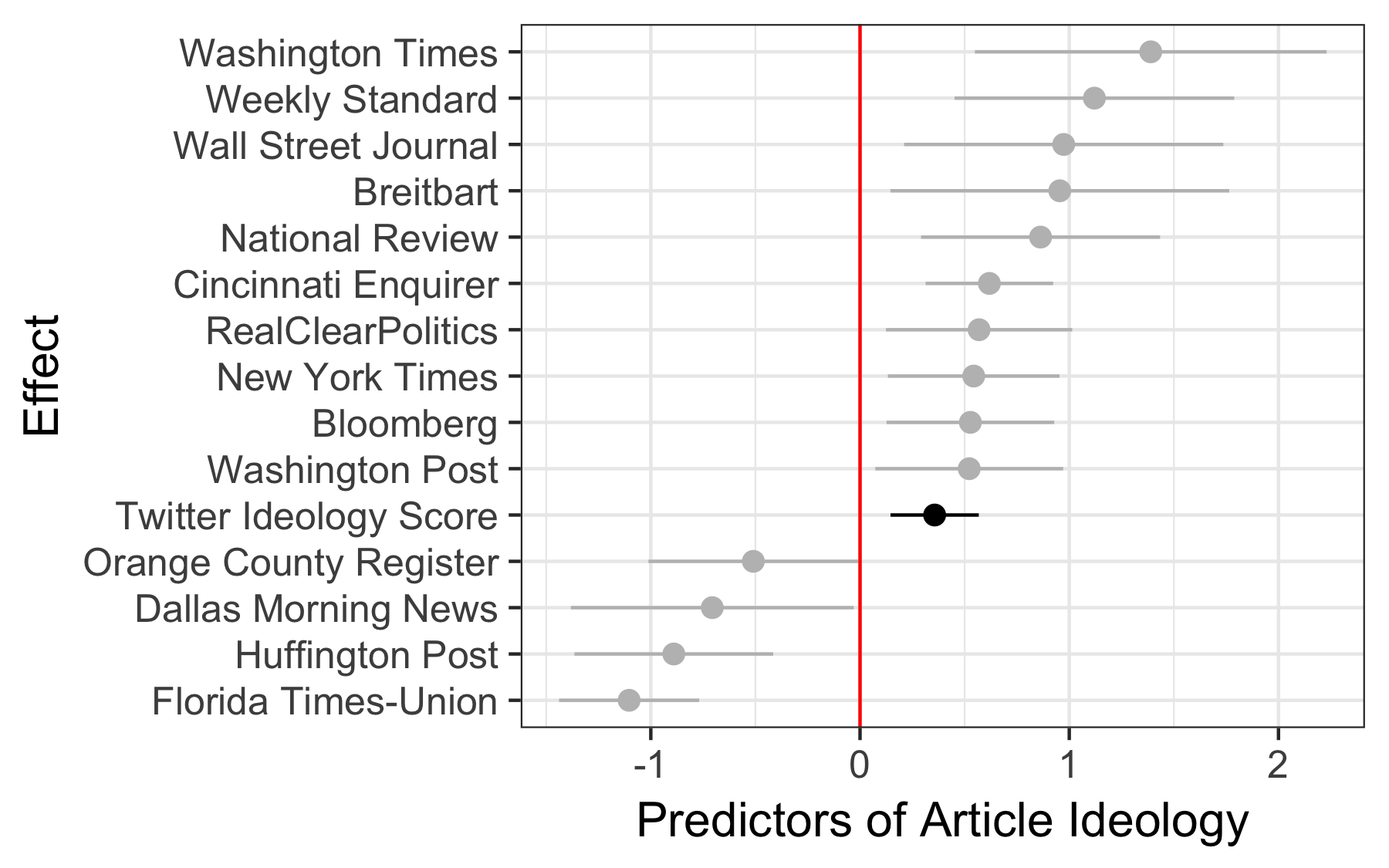}
	\caption{Regression coefficients for a linear model predicting textual ideology from Twitter ideology, controlling for outlet. Coefficients are shown with 95\% confidence intervals; only coefficients whose 95\% intervals do not cross zero are plotted}
	\label{fig:lm_output}
 \vspace{-1.3em}
\end{figure}

We now turn to the more direct question of how, controlling for biases induced by the outlet a journalist works for, the ideology of the content a journalist produces is correlated with the ideology of the accounts he or she follows on Twitter. To do so, we run a linear regression where the outcome variable is $s(j)$, and the independent variables are the outlet a reporter works for as well as the ideological leaning of the accounts the journalist follows. For the former set of variables, we use \emph{Politico} as the reference case. For the latter value, we standardized by subtracting the mean and dividing by two standard deviations for ease of interpretation \cite{gelman_scaling_2008}.

Figure~\ref{fig:lm_output} presents results for coefficients significant at p <.05, and shows that the ideology of a journalist's Twitter network is moderately but significantly correlated with the ideology of the content they produce, even when controlling for the outlet a journalist works for.  An increase in two standard deviations in the level of right-leaning content in a journalist's Twitter network is associated with an increase of .35 in production of right-leaning content ($s(j)$). In other words, a journalist whose twitter feed is mostly conservative is likely to produce content that is approximately one quarter of one standard deviation more conservative than a journalist whose feed is mostly liberal. 

\section{Discussion}
While we find a clear correlation between the ideology of the journalist's Twitter network and the ideology of his or her writing, a closer inspection of specific journalists yields important exceptions. For example, outliers such as David Sanger of the \emph{New York Times} focus heavily on national security and military affairs. While Sanger would \emph{not} be characterized as particularly partisan by most media observers, our analysis decisively places him amongst the most right-leaning content producers, even though his Twitter network is among the most left-leaning. Beyond Sanger, who is an extreme outlier, there are a substantial number of journalists whose Twitter networks lean left, but whose content focuses on right-leaning topics.  

Survey research continues to suggest that journalists as a whole are more likely to be left-leaning \cite{willnat_global_2013}.  The fact that journalists are generally concentrated in metropolitan areas, which tend to vote more liberally in elections, is plausibly a factor in many of their online social networks skewing in the same ideological direction. Yet the findings here give tentative evidence that professional codes of impartiality may remain strong in journalistic culture, despite elite criticisms, particularly from conservative commentators, that journalists are mostly liberals and therefore their professional output must be biased. The relationship between social networks and work output of journalists is complex and evolving, and the moderate correlation and obvious exceptions found point to that reality. 

Further, setting aside the specific nature of the correlations, the research inquiry itself here can render general applications both in the professional and academic fields of journalism. In considering their professional practices, it is highly useful---indeed, essential--- that journalists should reflect critically on the patterns of information they consume online and offline, and be more highly attuned to how bias and assumptions can slip into their work. By being more aware of what they are exposed to on Twitter, journalists might more objectively become aware of how their sense of reality is being shaped and how their understanding of the world is framed by sources they follow. In this regard, future research might employ survey methods to see how journalists' impressions of their own social networks, as well as their online sourcing and research routines, differ from the empirical realities.

\section{Conclusion}

The present work provides what is to our knowledge the first empirical study of journalists' online social networks and the connections with professional output. In an era where issues of political polarization and media bias are increasingly front-and-center, these connections between online social influence and the way public information is produced are of major importance. The chief finding presented is a significant, albeit moderate, correlation between the ideological character of a journalist's social network and ideological dimensions of his or her published output, even when controlling for the outlet for which the journalist writes. This result furnishes a crucial first step toward greater critical examination of emerging patterns of media bias. 

It is important to note, however, that the simplistic methods we use have important limitations.  More specifically, our approach to extracting ideological leanings of Twitter accounts could be extended to a more formally distantly supervised model based on politically active users, rather than simply adding these accounts to the matrix of journalists as we do here.  Our approach to extracting ideology from news articles should also be extended to more readily consider how the phrases discussed in news articles are ``spun'' or framed \cite{tsur_frame_2015,card_media_2015}.  Finally, while we provide informal validation of our methods, a more formal evaluation step is necessary if they are to be adopted for other purposes.


\bibliographystyle{ACM-Reference-Format}
\bibliography{sigproc} 

\end{document}